\documentclass{aa}
\usepackage{graphicx}
\usepackage[varg]{txfonts}

\usepackage{latexsym}
\usepackage{natbib}
\usepackage{amssymb}
\usepackage{amsmath}

\newcommand{\re}{$R_e$}
\newcommand{\Lsig}{$L=L_0\sigma^{\alpha}$}
\newcommand{\Lsiga}{$\log(L)-\log(\sigma)$}
\newcommand{\Lsigb}{$L=L'_0\sigma^{\beta}$}

\newcommand{\MR}{$\log(R_e)-\log(M^*)$}

\newcommand{\logre}{$\log(R_e)$}
\newcommand{\muem}{$\langle\mu\rangle_e$}

\newcommand{\IeRe}{$\log(\langle I\rangle_e)-\log(R_e)$}
\newcommand{\muere}{$\langle \mu\rangle_e-\log(R_e)$}
\newcommand{\FPR}{$\log(\sigma)-\log(\langle I\rangle_e)-\log(R_e)$}

\newcommand{\bfilt}{$\rm{B}$}
\newcommand{\vfilt}{$\rm{V}$}

\begin{document}

        \title{A tomography of the \IeRe\ plane}
        \subtitle{}




   \author{M. D'Onofrio \inst{1,2} \and
        C. Chiosi \inst{1}
}

\institute{Department of Physics and Astronomy, University of Padua,
        Vicolo Osservatorio 3, I-35122 Padua, Italy \\
        \email{mauro.donofrio@unipd.it}
        \and
        INAF -- Osservatorio Astronomico di Padova, Vicolo Osservatorio 5, I-35122 Padova, Italy \\
}

\date{Received November, 2020; Accepted xxx}

\abstract
{We present a re-analysis of the distribution of galaxies in the \IeRe\ plane under a new theoretical perspective.}
{{Using the data of the WINGS database and those of the Illustris simulation we will demonstrate that the origin of the observed distribution in this parameter space can be understood only by accepting a new interpretation of the \Lsiga\ relation.}}
{{We simulate the distribution of galaxies in the \IeRe\ plane starting from the new \Lsigb\ relation proposed by \cite{Donofrioetal2020} and we discuss the physical mechanisms that are hidden in this empirical law.}}
{The artificial distribution obtained assuming that $\beta$ spans either positive and negative values and that $L'_0$ changes with $\beta$, is perfectly superposed to the observational data, once it is postulated that the Zone of Exclusion (ZoE) is the limit of virialized and quenched objects. }
{{We have demonstrated that the distribution of galaxies in the \IeRe\ plane is not linked to the peculiar light profiles of the galaxies of different luminosity, but originate from the mass assembly history of galaxies, made of merging, star formation events, star evolution and  quenching of the stellar population. }}


\maketitle

\section{Introduction} \label{sec:intro}

\noindent
The $I_e -R_e$ relation, i.e. the distribution of galaxies in the parameter space formed by the effective radius (the radius enclosing half the total luminosity of a galaxy) and the surface brightness inside this radius, otherwise known  as \muere\ relation (when the units of $mag\, arcsec^{-2}$ are used instead of $L_\odot pc^{-2}$), was discovered in 1977 by Kormendy \citep{Kormendy1977}.
Today we know that  this space is a projection of the Fundamental Plane, i.e. the planar distribution observed for early-type galaxies  (ETGs) in the \FPR\ space \citep{Dressleretal1987,Djorgovski&Davis1987}. 
 
The  $I_e -R_e$ is the most easily accessible correlation of galaxies parameters for galaxies at low and high redshift. 
The first version of this relation, showed a linear correlation between \muem\ and \logre\ for ETGs with a slope $\sim3$ ($\sim-1.5$ in $I_e$ units). As soon as new data for faint ETGs and spiral objects became available, the observed distribution started to show an ample curvature, clearly separating  faint and dwarf objects with respect to bright ETGs.
This suggested the existence of two different populations of ETGs, the 'ordinary' and the 'bright', following two different trends in this space \cite{Capacciolietal1992}. The 'ordinary' family is bi-parametric ($L\propto I_e R_e^2)$, its members are fainter than $M_B \sim -19$ and their radii are smaller than $R_e \sim 3$ kpc. The 'bright' family is mono-parametric ($I_e$ depends only on $R_e$), it hosts only the brightest cluster galaxies (BCGs) and their members have radii larger than $R_e=3$ kpc. The spiral galaxies and their bulges belong to the 'ordinary' family and are not visible in the 'bright' sequence \citep{Donofrioetal2020}. 

Such curved distribution  has been used (among others correlations) to argue for distinct channels of formation for dwarfs and giants ETGs \citep[see e.g.][]{Capacciolietal1993,Kormendyetal2009,KormendyBender2012,Tolstoyetal2009,SomervilleDave2015,Kormendy2016}. 
Many authors believe that there are two distinct kinds of ETGs, whose properties differ mainly for the different history of merging events, in particular for the characteristics of the last major mergers, wet or dry, according to whether cold gas dissipation and starbursts occurred or not.

The existence of two physically distinct families of ETGs is at the center of an ample debate even  today.
Other researches, that did not use the effective half light radius parameter, advocated for a continuity among the ETG population \citep{Caldwell1983,BinggeliSandageTarenghi84,Bothunetal1986,CaldwellBothun1987}.
\cite{Graham2019} in particular  explored a range of alternative radii, including those where the projected intensity  drops by a fixed percentage, showing that the transition at $M_B\sim-19$ mag is likely artificial and does not mark a boundary between two different types of ETGs.

The shape of the light profiles of ETGs has been also used to claim a difference between dwarfs and ordinary ETGs: dwarfs have in general exponential light profiles similar to that of  late-type galaxies (LTGs)), while ordinary ETGs have $R^{1/n}$ profiles, with $n\geq3$. However, exponential light profiles are reproduced by the S\'ersic law when $n=1$. According to \cite{Graham2019}  the curved distribution of ETGs in the $I_e -R_e$ space, is likely associated to the continuous change of the S\'ersic index $n$ with the absolute magnitude (the $M_B - n$ relation found by \cite{Caonetal1993} and \cite{Donofrioetal1994}). In the same vein \cite{GrahamGuzman2003} argued that the only magnitude of importance in the $I_e -R_e$ plane is at $M_B=-20.5$ mag, where it is visible a division between the spheroidal components with S\'ersic profile and those with core-S\'ersic profile. This magnitude corresponds to a mass of $\sim2 \times 10^{11} M_\odot$.

There are indeed two linear scaling relations involving the structural parameters of ETGs: the  $M_B - \mu_0$ relation between total luminosity and central surface brightness  and the $M_B - n$ relation between total luminosity and S\'ersic index $n$. These relations do not show evident signs of curvature. The first one is a re-statement of the concentration classes introduced by \cite{Morgan1958}, later quantified by the concentration index $C$ \citep{Fraser1972,BinggeliSandageTarenghi84,Kent1985,Ichikawaetal1986}. The second is a consequence of the first, being the S\'ersic parameter a measure of the radial concentration of galaxy light. Further examples of the $M_B - n$ diagram have been derived by \cite{YoungCurrie1994}, \cite{Grahametal1996}, \cite{Jerjenetal2000}, \cite{Ferrareseetal2006} and \cite{Kormendyetal2009}. 

The lack of curvature in these diagrams does not support the view of different formation mechanisms at work for the formation of ETGs. The same can be said for the \Lsig\ Faber-Jackson (FJ) relation \citep{FaberJackson1976}, where faint and bright ETGs follow the same trend, with only few cases of small deviations (see below). However, 
a deep analysis from \cite{Nigoche-Netroetal2011} concluded that the intrinsic dispersion of the FJ relation depends on the history of galaxies, i.e.  on the number and nature of transformations that have affected the galaxies along their  lifetimes (collapse, accretion, interaction and merging).

Another interesting feature of the $I_e - R_e$ diagram, is the presence of a zone of exclusion (ZoE), a region strictly  avoided by galaxies. The distribution of galaxies appears limited in the maximum surface brightness at each $R_e$. The slope of this line of avoidance is $-1$ in these units, i.e.  the slope predicted by the Virial Theorem \citep{Donofrioetal2020}.
The existence of the ZoE was first noted by \cite{Benderetal1992} and \cite{BBF} using the $k$-space version of the FP. They described the ZoE with the equation $k_1+k_2\leq7.8$. In the $k$-space
the dynamically hot stellar systems appear segregated, with a major sequence formed by luminous ellipticals, bulges and some compact elliptical. A second sequence is formed by dwarf ellipticals and dwarf spheroidals. The $k$-space was used to infer the main physical properties of galaxies, looking at the role played by merging, dissipation, tidal stripping and winds. We will dedicate a separate work to the analysis of this space (Chiosi \& D'Onofrio, in preparation). 

In this paper we will address the origin of the curved distribution observed in the  $I_e -R_e$ plane, starting from a different perspective, i.e. considering the role played by Faber-Jackson relation in its different formulation proposed by \cite{Donofrioetal2020}, i.e. through the \Lsigb\ relation, where $L'_0$ and $\beta$ are variable factors depending on the mass assembly history of galaxies. 

In Sec. \ref{sec:1} we describe the data used in our plots, then we present in Sec. \ref{sec:2} our simulation of the $I_e -R_e$ plane obtained using different $\beta$ and $L_0$.  Finallly in Sec. \ref{sec:3} we discuss the origin of the observed distribution and present our conclusions.

In the paper we used the data of the WINGS database (see below) that have been derived assuming the standard values of the $\Lambda$-CDM cosmology \citep{Hinshaw_etal_2013}:
$\Omega_m = 0.2726, \Omega_{\Lambda}= 0.7274, \Omega_b = 0.0456, \sigma_8 = 0.809, n_s = 0.963, H_0 = 70.4\, km\, s^{-1}\, Mpc^{-1}$.

\section{The Sample} \label{sec:1}
The observational data used in this study are extracted from the WINGS and Omega-WINGS database
\citep{Fasano2006,Varela2009,Cava2009,Valentinuzzi2009,Moretti2014,Donofrio2014,Gullieuszik2015,Morettietal2017,
Cariddietal2018,Bivianoetal2017}.

The WINGS and Omega-WINGS datasets are the largest and most complete data samples for galaxies in nearby clusters ($0<z<0.07$). The database includes galaxy magnitudes, morphological types, effective radii, effective surface brightness, stellar velocity dispersion, star formation rates and many other useful measurements obtained by the WINGS team.

The WINGS optical photometric catalog is 90\% complete at \vfilt\ $\sim 21.7$ \citep{Varela2009}. The
database includes respectively 393013 galaxies in the \vfilt\ band and 391983 in the \bfilt\ band. The cluster
outskirts were mapped with the Omega-WINGS photometric survey at the VST telescope \citep{Gullieuszik2015} covering 57 out of 76 clusters.

The  data extracted from the WINGS database \citep{Moretti2014} are: \\
1. the  aperture corrected velocity dispersions of 1729 ETGs, measured by the Sloan Digital Sky Survey (SDSS) and by the National Optical Astronomical Observatory (NOAO) survey, already used by \cite{Donofrioetal2008} to infer the properties of the FP; \\
2. a set of new measured velocity dispersions derived by \cite{Bettoni2016}; \\
3. the effective radii and surface brightness in the V-band of 34982 galaxies, either ETGs and LTGs members and non-members of our clusters, derived by
\cite{Donofrio2014} through the software GASPHOT \citep{Pignatelli}; \\
4. the total luminosities and distances derived from the redshifts measured by
\citep{Cava2009,Morettietal2017}.

In addition to real data we used the ample database of artificial galaxies simulated by the Illustris simulation
\citep[][to whom we refer for all details]{Vogel2014,Genel_etal_2014,Nelsonetal2015}.
A full description of this dataset is available in \cite{Cariddietal2018} and \cite{Donofrio2019}.
In brief we used the run with full-physics (with both baryonic and dark matter) having the highest degree of
resolution, i.e. Illustris-1 \citep[see Table 1 of][]{Vogel2014}, extracting in particular the \vfilt-band
photometry, the mass, and the half-mass radii of the stellar particles (i.e., the integrated stellar populations), as well as the comoving coordinates $(x',y',z')$.

The projected light and mass profiles using the $z'=0$ plane as reference plane were studied in the paper of \cite{Donofrio2019}. Starting from the \vfilt\ magnitudes and
positions of the stellar particles, we computed the effective radius \re\ and effective surface brightness
\muem, the radial surface brightness profiles, the best-fit S\'ersic index and the
line-of-sight velocity dispersion $\sigma$ following
\citet{Zahid_etal_2018}. 

Furthermore, in order to follow the evolution of the galaxies back in time, we extracted from the Illustris database the stellar mass, the \vfilt\ luminosity, the half-mass radius, the velocity dispersion, and the SFR for the whole set of galaxies (with mass $\log(M^*)\geq9$ at $z=0$) in the selected clusters at redshift
$z=0$, $z=0.2$, $z=1$, $z=1.6$, $z=2.2$, $z=3$, and $z=4$. With these data we  were able to follow the progenitors of each object across
the epochs and compare observations with simulations up to redshift $z=4$.

\section{The $I_e -R_e$ plane}\label{sec:2}

\begin{figure}
	\centering
	        \includegraphics[width=0.45\textwidth]{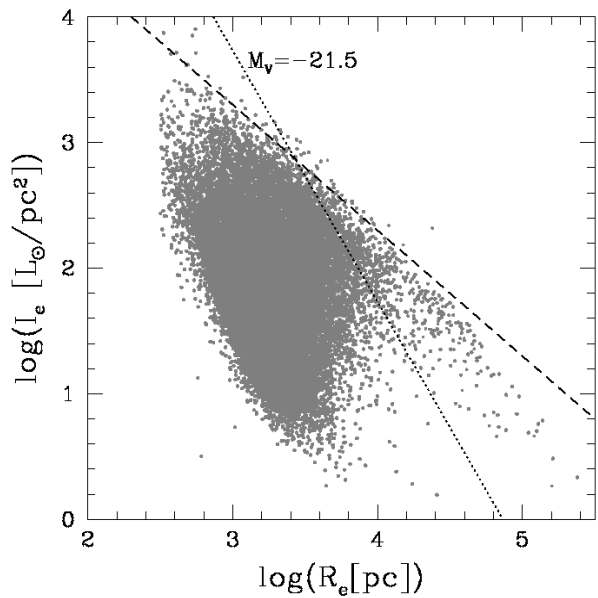}
	\caption{Distribution of galaxies in the $I_e - R_e$ plane. The gray dots mark galaxies of all morphological types. The dashed line marks the ZoE, i.e. the line with slope $-1$ that is predicted for virialized and quenched objects. The dotted line marks the locus of constant luminosity for $M_V=-21.5$. }
	\label{fig:1}
\end{figure}

The distribution of galaxies in the $I_e - R_e$ plane for more than 30000 galaxies of all morphological types is shown in Figure \ref{fig:1}.

The features of interest in such distribution are:\\
1. the tail at large effective radii ($R_e\geq4$ dex in pc units) for bright galaxies ($M_V\leq-21.5$);\\
2. the cloud of 'ordinary' galaxies with maximum radii of $\sim3-5$ kpc;\\
3. the sharp boundary due to the ZoE, i.e. the region avoided by galaxies of any type;\\
4. the lower limit in magnitude at $M_V\sim-15.5$, providing the maximum performance of the WINGS survey
in detecting faint objects.

The cloud of 'ordinary' galaxies and the tail of 'bright' galaxies defined by \cite{Capacciolietal1992} is clearly visible. As we mentioned in the introduction the origin of such dichotomy is still debated.

Here we want to propose a new way of looking at this distribution. The starting point in this case is the Faber-Jackson relation \Lsig\ \citep{FaberJackson1976}, i.e. the log linear relation observed between total luminosity and central velocity dispersion.

\begin{figure}
	\centering
	\includegraphics[width=0.45\textwidth]{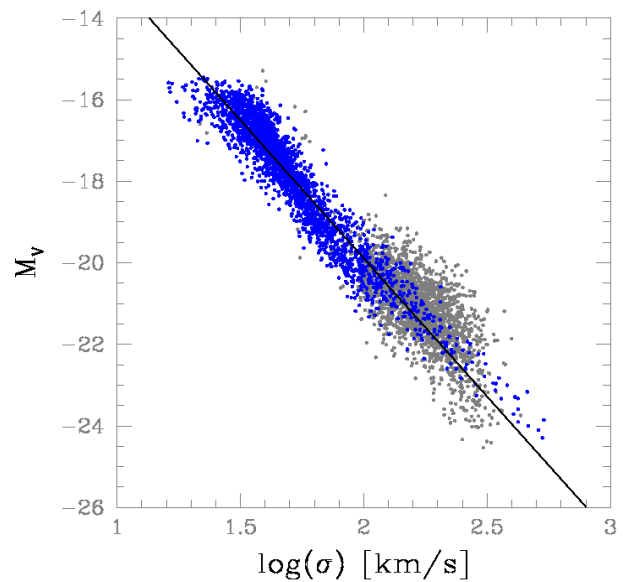}
	\caption{The FJ relation for the WINGS galaxies (gray dots) and the Illustris simulation (blue dots).}
	\label{fig:2}
\end{figure}

We can see the FJ relation in Fig. \ref{fig:2}, using $M_V$ instead of $\log(L)$ (this permits to compare the lower limits of the photometric and spectroscopic databases, close to $M_V=-16$).
The plotted data are those of the WINGS and Illustris databases. 
The observational data have approximately 1800 measurements of $\sigma$ for ETGs (gray dots). The Illustris database contains instead $\sim 2400$ objects.

We can see in the Figure that the two distributions are not different each other. In particular there are no evidences of curvature of the relation up to the faintest luminosities.
The fitted relation, marked by the black line, is:

\begin{equation}
M_V=-6.77(\pm0.05)\log(\sigma)-6.34(\pm0.05)
\label{eq1}
\end{equation}
\noindent
In $\log(L)$ units the slope is instead $\sim2.7$. The scatter is $\sim0.3$.

Now, we re-write the relation in $L$ units, recalling the definition of mean effective surface brightness $I_e$:

\begin{equation}
L=2\pi I_e Re^2=L_0 \sigma^{\alpha}.
\label{eq1a}
\end{equation}
\noindent
$\alpha$ was $\sim4$ in the first version of the FJ relation, while today it ranges from 2.7 to 3.5, according to the method and the data sample used to  fit the data \citep[see e.g.][]{Donofrioetal2020}. The relation seems to suggest that the zero-point $L_0$  is approximately constant for all types of ETGs of any luminosity and equal to $\sim 3.9$ with a scatter of $\sim 0.3$.

Passing to log units and solving for $I_e$ we get:

\begin{equation}
\log(I_e)=-\log(2\pi)-2\log( Re)+\log(L_0)+\alpha\log(\sigma)
\label{eq2}
\end{equation}

\begin{figure}
	\centering
	\includegraphics[width=0.45\textwidth]{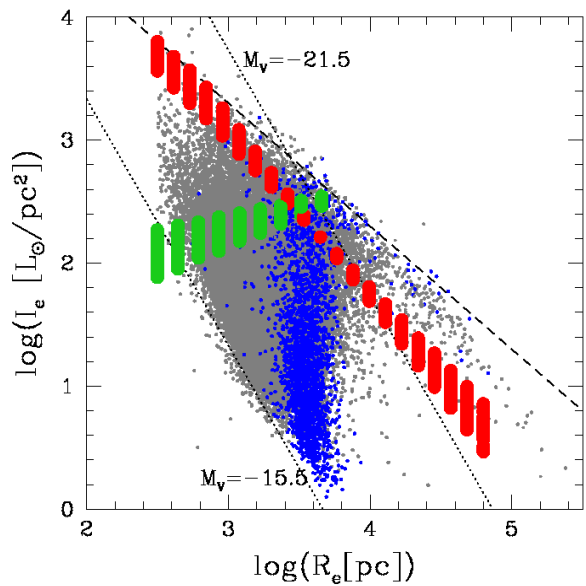}
	\caption{The same $I_e - R_e$ plane of Fig. \ref{fig:1} with superposed in green and red colors the expected distribution of $I_e$ obtained from eq. \ref{eq2} and eq. \ref{eq3} respectively (see text).}
	\label{fig:3}
\end{figure}

\begin{figure}
	\centering
	\includegraphics[width=0.45\textwidth]{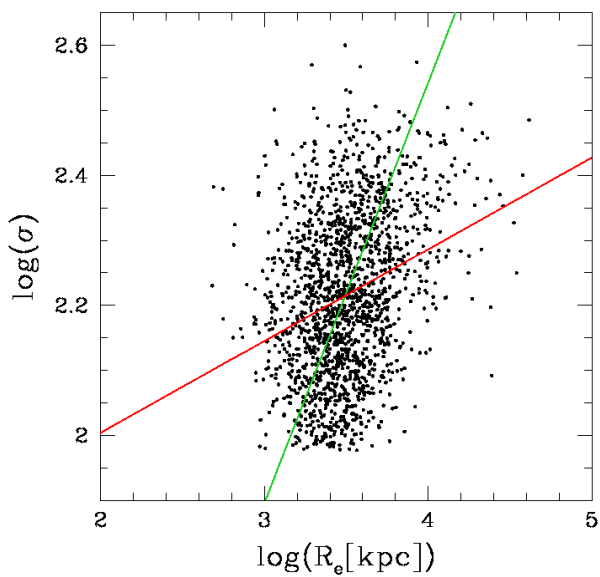}
	\caption{The $R_e - \sigma$ relation in the data of the WINGS database. The black dots mark the observational data. The green line is the fit obtained by SLOPES through the bilinear least square fitting procedure. The red line marks instead the standard least square fit.}
	\label{fig:4}
\end{figure}

Figure \ref{fig:3} shows again the $I_e -R_e$ plane with superposed the expected values of $\log(I_e)$ obtained from eq. \ref{eq2} and eq. \ref{eq3}, by varying $R_e$ in the observed interval of possible values and assuming for $\sigma$ the values expected from the fit of the $R_e - \sigma$ relation, derived using the data of the WINGS survey. The fit of the data is visible in Fig. \ref{fig:4} and was done with the SLOPES software \citep[][]{Feigelson1992}. The bisector fit (in green color in the figure) gives:

\begin{equation}
\log(\sigma)=0.64(\pm0.02)\log( Re[pc])-0.046(\pm0.071)
\label{eq2a}
\end{equation}
\noindent
with an rms scatter of 0.11, a correlation coefficient of 0.3 and  a significance of $8.6e-37$.
The standard least square fit (in red color) gives instead:

\begin{equation}
\log(\sigma)=0.14(\pm0.01)\log( Re[pc])+1.722(\pm0.037)
\label{eq3}
\end{equation}

It is apparent that the distribution in the $I_e -R_e$ plane depends critically on the assumed \Lsig\ relation and on the $R_e - \sigma$ relation. The slope of the simulated distribution depends critically on the slope of the $R_e - \sigma$ relation, while the scatter in the distribution depends on the permitted variation of $L_0$. 

We conclude that the standard FJ relation is incompatible with the observed distribution in the $I_e - R_e$ plane.

Why is  this possible? What element enters in the $I_e - R_e$ relation that is not present in the \Lsig\ relation?
According to \cite{Donofrio2017} the problem resides in a mis-understanding of the \Lsig\ relation.
The linear relation that we see in the FJ is due to the fact that we are looking at the relation between mass and velocity dispersion, i.e. to a modified version of the virial theorem, written using $L$ instead of $M$, while the true relation between $L$ and $\sigma$ may  has a different origin. We make the \textit{ansatz} that the relation between luminosity and velocity dispersion for each galaxy keeps the formal dependence of the  FJ relation but with different exponent and proportionality factor that can vary from galaxy to galaxy:

\begin{equation}
L = L'_0 \sigma^{\beta},
\label{eq4}
\end{equation}

\noindent
with $L$ in solar luminosities. Here $L'_0$ is  a proportionality factor that strongly depends on the star formation history of each galaxy, and the exponent $\beta$  reflects the peculiar motion of each object in the $\log(L) - \log(\sigma)$ plane across the cosmic epochs due to merging and star formation events. In this case the relation is different for each galaxy, being $L'_0$ and $\beta$ different from galaxy to galaxy. These two variables must be intimately  connected each other in order to keep small the scatter of the observed \Lsiga\ relation. 

In the following we intend   to  explore the possibility that the true relation determining the observed dichotomy in the $I_e - R_e$ plane is this \Lsigb\ relation. In other words we want to study the relation:

\begin{equation}
\log(I_e)=-\log(2\pi)-2\log( Re)+\log(L'_0)+\beta\log(\sigma)
\label{eq5}
\end{equation}

The main problem in this relation is that we do not know the present values of $L'_0$ and $\beta$ for each galaxy. This is why we need the data of the Illustris simulation.

Through numerical simulations we can in fact demonstrate that $\beta$ and $L'_0$ are subject to variations from object to object and across the cosmic epochs. In particular the slope $\beta$ turns out to have a spectrum of values ranging from large negative to large positive \citep[see][]{Donofrioetal2020}.

\begin{figure}
	\centering
	\includegraphics[width=0.45\textwidth]{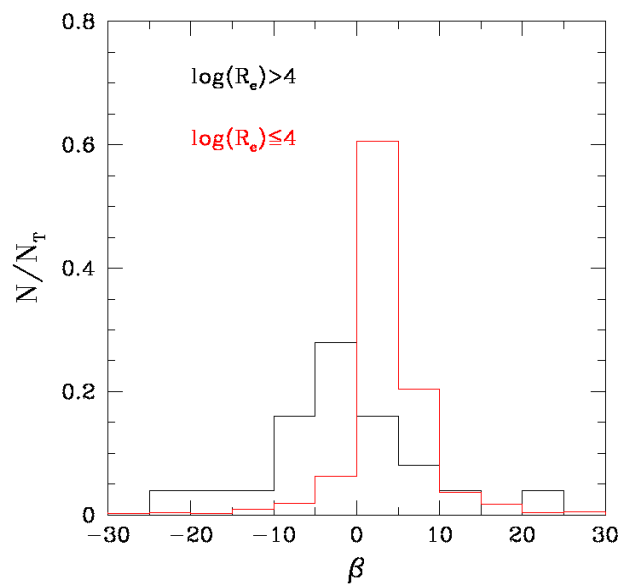}
	\caption{Histogram of the values of $\beta$ derived from the Illustris simulation. The black (red) histogram is done for objects that have $\log(R_e)>4$ ($\log R_e\leq4$).}
	\label{fig:5}
\end{figure}

Here we have re-derived the possible value of $\beta$ by considering the variation in luminosity $\Delta\log(L)$ and central velocity dispersion $\Delta\log(\sigma)$ between the different redshift epochs of the Illustris simulation. 

Figure \ref{fig:5} shows for example the histogram of the values of $\beta$ obtained considering the two redshift epochs at $z=1$ and $z=0$.  The values of $\beta$ are obtained from the ratio $\Delta\log(L)/\Delta\log(\sigma)$.
This ratio figures out the direction of motion of a galaxy in the \Lsiga\ plane between the two cosmic epochs. We can see clearly that the values of $\beta$
are either positive and negative. In the figure we plotted with a black line the histogram of the galaxies that have large $R_e$ ($>4$ dex) and with a red line those with $R_e\leq4$. Note how the big galaxies have preferentially negative values of $\beta$; this means that  their luminosity  decreased at nearly constant $\sigma$ in that time interval. They are likely objects undergoing a quenching phase.

\begin{figure}
	\centering
	\includegraphics[width=0.45\textwidth]{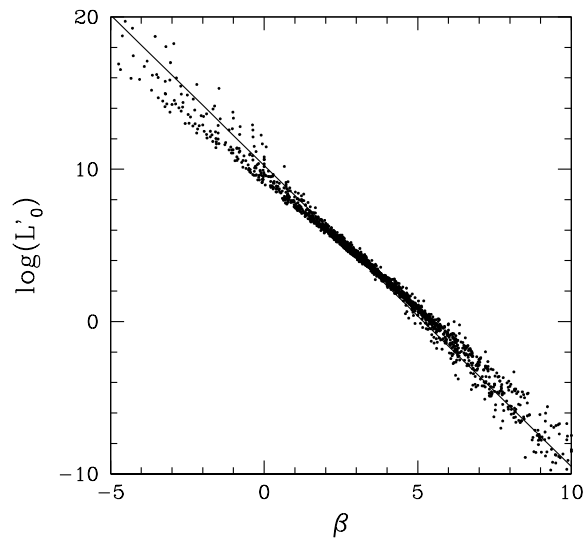}
	\caption{The relation between $\beta$ and $L'_0$ derived from the Illustris simulation. We plotted only the central interval with the most frequent values of $\beta$.}
	\label{fig:6}
\end{figure}

Once $\beta$ is known, we must  derive $L'_0$. This is obtained by using
Eq. \ref{eq4} that gives us the values of $L'_0$ for each galaxy (because the values of $L$ and $\sigma$ are know at each redshift epoch). Using the Illustris data we find the linear relation between  $L'_0$ (in log units) and $\beta$ shown in Fig. \ref{fig:6}, that is given by:  

 \begin{equation}
 \log(L'_0)=-1.976\,\beta+10.246
 \label{eq6}
 \end{equation}
\noindent
with an rms $\sim5$ and a c.c. $\sim0.99$.

This behavior is expected because the \Lsig\ relation is tight and with a small scatter.
The same procedure can be used for all the different redshift epochs available in our Illustris database. Table \ref{tab:1} shows the slopes, intercepts, rms scatters and correlation coefficients for the relations between $L'_0$ and $\beta$ obtained considering all redshift intervals.

\begin{table}
	\begin{center}
		\caption{Slopes, intercepts, rms scatter and c.c. of the relations between $\beta$ and $L'_0$ extracted from simulations.}
	\begin{tabular}{rrrrl}
		\hline
     slope  & intercept &    rms &   c.c. & z interval \\
\hline\hline
-1.738    &    8.960  &   7.405  &   -0.9942 & z=4-->z=3 \\
-1.775    &    9.429  &   6.171  &   -0.9965 & z=3-->z=2.2 \\
-1.763    &    9.660  &   5.755  &   -0.9963 & z=2.2-->z=1.6 \\
-1.881    &  10.070  &   6.962  &   -0.9928 & z=1.6-->z=1 \\
-1.916    &  10.170  &   6.149  &   -0.9932 & z=1-->z=0.6 \\
-1.990    &  10.420  &   5.835  &   -0.9923 & z=0.6-->z=0.2 \\
-1.789    &    9.684  &   5.177  &   -0.9928 & z=0.2-->z=0 \\
\hline
\label{tab:1}
\end{tabular}
\end{center}
\end{table}

We can see that the slope and intercept smoothly vary across the redshift epochs. We can then explore the effects of varying $\beta$ and $L'_0$ in eq. \ref{eq5}.
By inserting eq. \ref{eq6} and eqs. \ref{eq2} (or \ref{eq3}) in eq. \ref{eq5} we can observe what happen
to $I_e$ when we vary the effective radius $R_e$ in the observed range.

\begin{figure}
	\centering
	\includegraphics[width=0.45\textwidth]{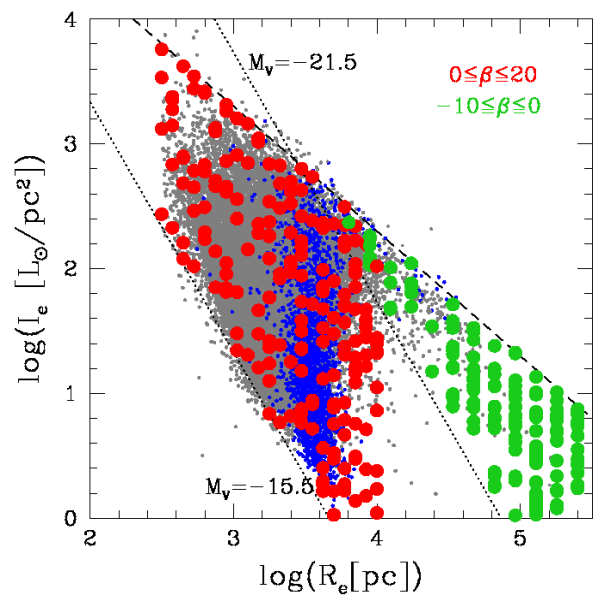}
	\caption{The $I_e - R_e$ plane of Fig. \ref{fig:1} with superposed in blue color the data of the Illustris simulation. Red and green dots mark the artificial data obtained from the combination of eqs \ref{eq3}, \ref{eq5} and \ref{eq6} according the variation of $\beta$. The artificial data above the ZoE have not been plotted as well as those below the limiting surface brightness of the observational data.}
	\label{fig:7}
\end{figure}

Figure \ref{fig:7} shows that when the new \Lsigb\ FJ relation is inserted in eq. \ref{eq1} the $I_e - R_e$ plane is well reproduced, i.e. the real and artificial distributions are almost exactly superposed. 

The only assumption done in addition to that of variable values of $\beta$ and $L'_0$ is that no galaxies are admitted in the region of the ZoE. We have not plotted them in the diagram. We also do not plot the points that are below the limiting surface brightness of the WINGS survey. 

It is also important to keep in mind that by varying $\beta$ and $L'_0$ it is possible to observe galaxies with $\beta<0$ among the 'ordinary' family. However in Fig. \ref{fig:7} we plotted only those with $\beta<0$ belonging to the 'bright' family.

In conclusion we can note the following:
1. The data with negative values of $\beta$ do fill the region of 'bright' galaxies.
2. The data with positive values of $\beta$ are instead in the region of the 'ordinary' galaxies.

We stress here that the distribution visible in Fig. \ref{fig:7} is obtained by varying of small amounts the values of the zero-point and slope of the $L'_0 - \beta$ relation. The distribution of galaxies in this space critically depends on: a) the assumed relation between $R_e$ and $\sigma$; b) the values adopted for the zero-point and slope of the $L'_0 - \beta$ relation. The density of plotted points depends instead on the maximum interval of values of $\beta$, either positive or negative, that is adopted.

In addition we can observe that:
1. the $I_e - R_e$ distribution is not in contrast with the FJ relation, if the 'correct' relation between luminosity and velocity dispersion is used (i.e. the \Lsigb\ relation);\\
2. the negative values of $\beta$ are those permitted to objects that are today close to the virial equilibrium and in a passive state of evolution, i.e. objects whose luminosity is now decreasing at nearly constant $\sigma$;\\
3. the shape of the distribution does not depend on the shape of the light profiles of the galaxies, but only to the evolution of $L$ and $\sigma$. Both depend on the merging and star formation events experienced by galaxies, i.e. on the mass assembled during these events. Star evolution instead affects only luminosity, producing a natural quenching of the star emission as time goes by.
 
 \section{Discussion and conclusions}\label{sec:3}
 
Using the empirical relation \Lsigb, we have demonstrated that the $I_e - R_e$ distribution can be deduced by accepting the idea that luminosity and velocity dispersion are mutally correlated variables.
The \Lsigb\ relation hides the complex relationship existing between the baryon and DM components
and the history of mass accretion and stellar evolution experienced by each stellar system. This relation is independent of the virial theorem. 

The idea behind this new perspective is that the total luminosity of galaxies is essentially the result of the mass assembly, star formation history  and star's evolution. Luminosity is therefore a non monotonic function of such variable factors. In 1973 \cite{Brosche1973} first suggested a failing of the simple star formation (SF) law of \cite{Schmidt1959}, based only on the gas density $\rho$, favoring a scenario in which the SF is a function $\sim f(\rho v^{\beta})$, where $v$ is the velocity of stars and $\beta$ is a variable factor for most of the galaxies. Stars born in large gas aggregates have a characteristic velocity that depends on the physical condition of the galaxy during the SF event (collapse, shock, merging, etc.). For this reason the global SF might keep memory of the velocity of this gas.

\begin{figure*}
	\centering
	\includegraphics[width=1\textwidth]{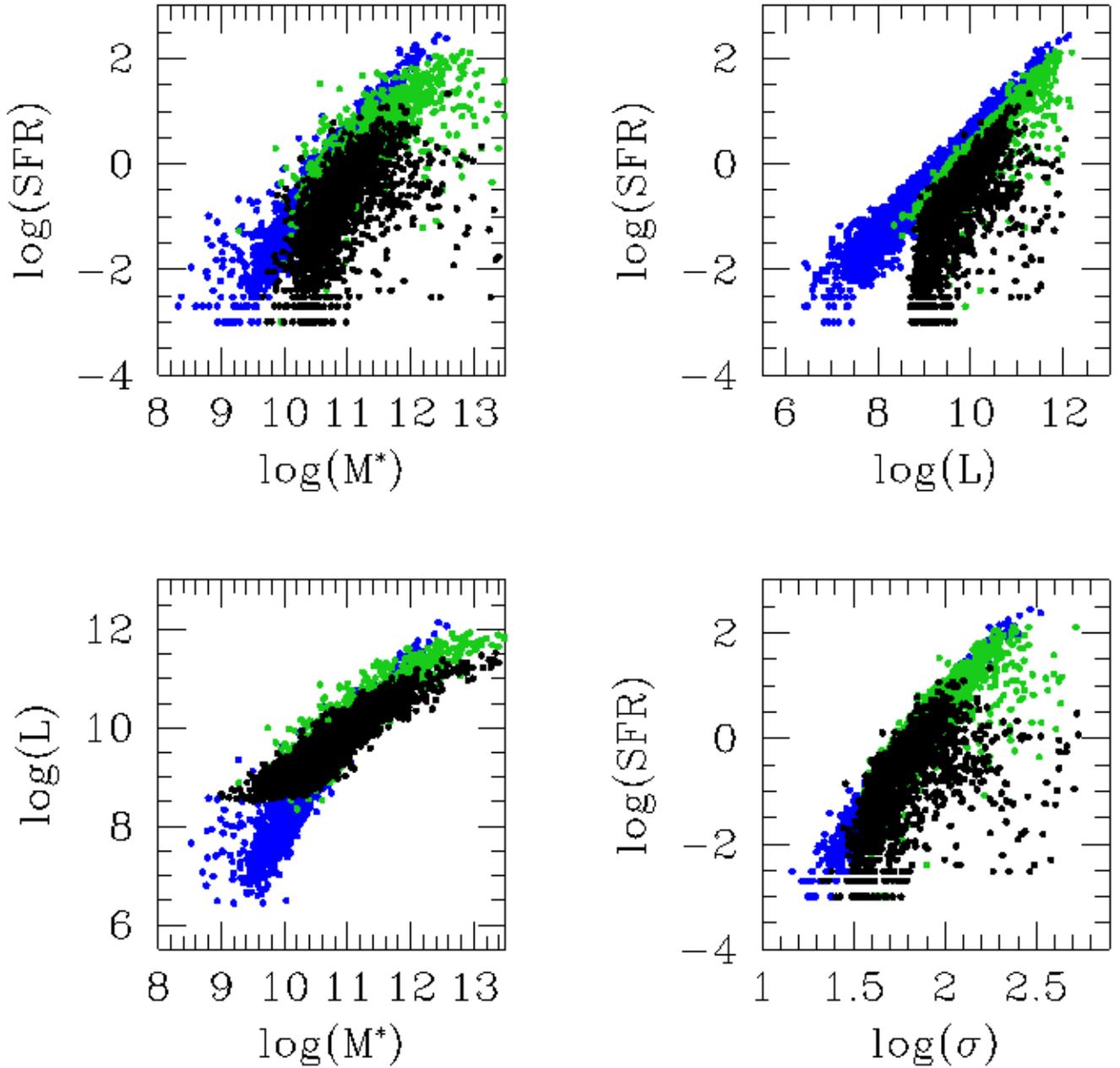}
	\caption{Upper left diagram: the $M^* - SFR$ relation at three different redshift epochs: $z=4$ (blue dots), $z=1$ (green dots) and $z=0$ black dots. Upper right diagram: the $L - SFR$ relation. Lower left diagram: the $M^* - L$ relation. Lower right diagram: the $\sigma - SFR$ relation.}
	\label{fig:8}
\end{figure*}

Here, by exploiting the data of the Illustris simulation, we can look at the properties  of galaxies at different redshift epochs. Figure \ref{fig:8} shows for example four panels where we have compared the following parameters: the stellar mass $M^*$, the total luminosity $L$, the SFR and the stellar velocity dispersion. The data are those belonging to three different redshift epochs, respectively at $z=4$ (blue dots), $z=1$ (green dots) and $z=0$ (black dots). The upper left diagram shows the $M^* - SFR$ plane. It apparent that at $z=0$ many galaxies are going to shut down the SFR. The process is visible event at $z=1$, while is absent at $z=4$. The same trend is visible in the $L - SFR$ plane. The birghtest galaxies in particular are those where the SFR is drastically quenched. Notably the $M^* - L$ plane shows that the galaxies follow a progressively tilted relation, with quite similar scatter, much steeper at $z=4$ than at $z=0$.

The observed trends seems to suggest that the formation of the bright family tails in the \IeRe\ (and \MR\ planes) might be associated to the quenching process, i.e. to the shut down of the SF. We have already seen that the tails in these scaling relations is well reproduced by Illustris \citep{Donofrioetal2020}. The tails are formed by objects with mass higher than $10^{10} M_{\odot}$ after the cosmic epoch $z=2$.

Here we want to better analyze what processes might be originating the distribution of galaxies in the $I_e - R_e$ plane.
In \cite{Donofrioetal2020} we speculated that dry mergers could be responsible for the tail of 'bright' galaxies. The merging of stars without gas might inflate the stellar systems because the energy could not be dissipated. The absence of SF seems supported by the fact that  the majority of galaxies in the tail are red.

The observed tails in the \IeRe\ (and \MR\ planes) seem also connected with the existence of the ZoE. When a galaxy reaches the passive state it can also fully relax and become virialized.
The ZoE could therefore be a sort of universal limit established by the condition of full virialization and passiveness.
The ZoE indicates that an object of a given mass can never have a radius smaller than that achieved when
it reaches the undisturbed virialization and passive state.
Since no system can cross the ZoE, this line appears as a physical barrier in the \IeRe\ (and \MR) plane. Only the systems that have reached a full virialization and are today evolving in a pure passive way can be
distributed close to the tail. When a 'bright' galaxy is in this state of evolution $\beta$ is negative and the galaxy can reach at maximum the line slope predicted by the virial theorem.  This occurs only for the massive galaxies that are today poorly affected by minor mergers (major mergers are today very rare). These systems are the closest to the condition of full virialization. They are also passive
objects since their star formation quenched long time ago. Their luminosity therefore decreases at nearly constant $\sigma$.

For the objects of the 'ordinary' family the virial
equilibrium is very unstable since merging and stripping events and episodes of star formation rapidly move
the galaxies toward a new condition of virial equilibrium. These systems are not passive yet,  and are therefore far from the ZoE.

\begin{figure*}
	\centering
	\includegraphics[width=0.4\textwidth]{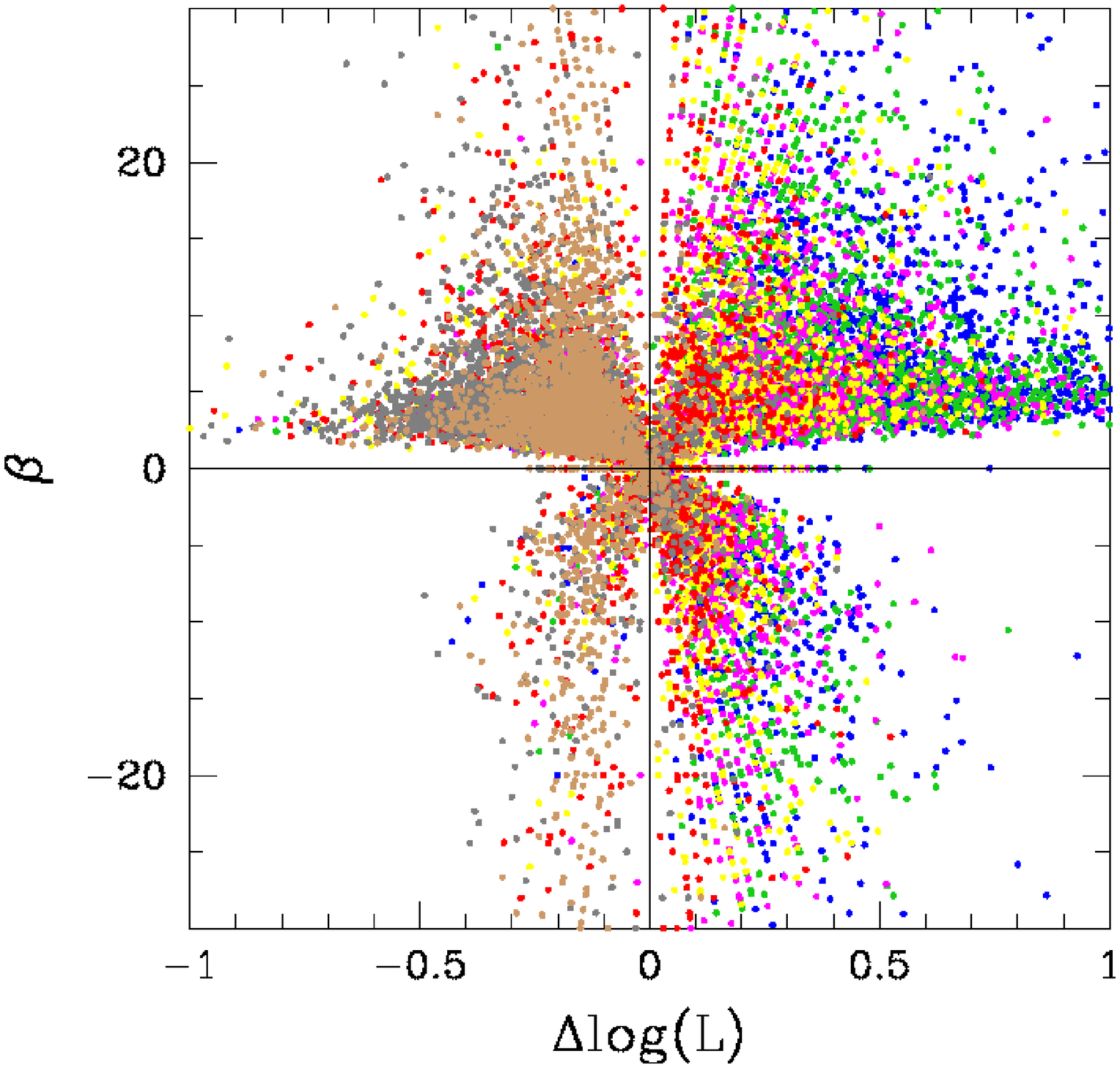}
	\includegraphics[width=0.4\textwidth]{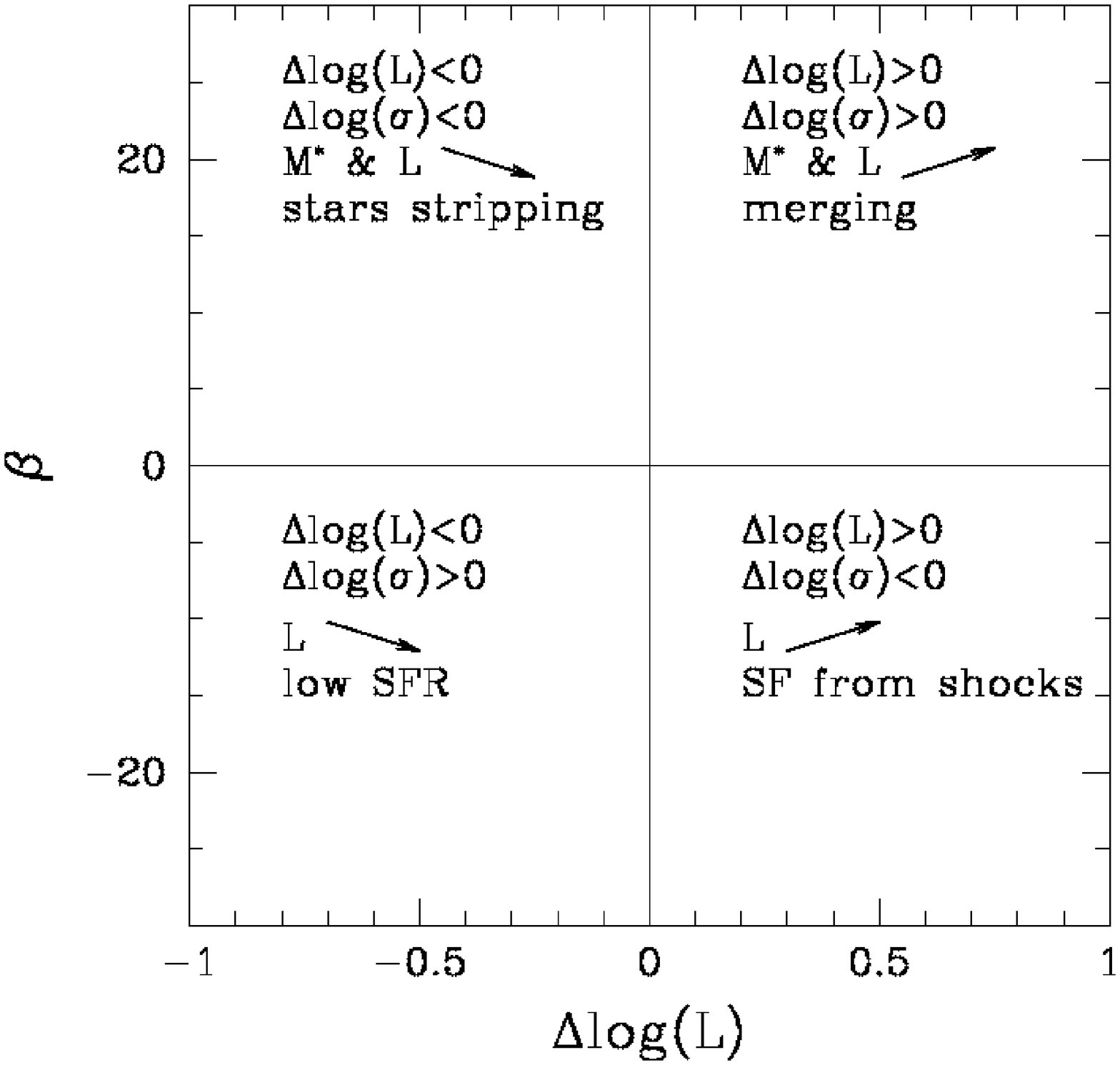}
	\caption{Left  diagram: the $\Delta\log(L) - \beta$ plane. The different redshift epochs are marked by dots of different colors: the most remote one, starting at $z=4$ and going to $z=3$ (blue dots), is followed by that from $z=3$ to $z=2.2$ (green dots), that from $z=2.2$ to $z=1.6$ (magenta dots), that from $z=1.6$ to $z=1$ (yellow dots), that from $z=1$ to $z=0.6$ (red dots), that from $z=0.6$ to $z=0.2$ (gray dots), and that from $z=0.2$ to $z=0$ (brown dots). Right diagram: a sketch of the same plane showing the variation of mass, luminosity, and velocity dispersion associated to each position of a galaxy in this diagram.}
	\label{fig:9}
\end{figure*}

In order to better understand what is going on in the $I_e - R_e$ and \Lsiga\ planes we decided to use the values of $\beta$ at each redshift interval (see Tab. \ref{tab:1} obtained using the Illustris data) and compare them with the variation in luminosity predicted by the simulation between the same redshift epochs $\Delta\log(L)$. Figure \ref{fig:9} shows in the left panel the $\Delta\log(L)-\beta$ plane for all redshift epochs.

The left panel of such figure shows that the galaxy transformation occured at nearby epochs, marked with gray and brown dots, are preferentially located in the left part of the plot, where $\Delta\log(L)<0$. On the other hand, those at high redshift are preferentially observed where $\Delta\log(L)>0$. Since $\beta$ is given by the ratio $\Delta\log(L)/\Delta\log(\sigma)$, the right panel of Fig. \ref{fig:9} gives a sketch of the properties of galaxies located in the different quadrants of this diagram. The galaxies that have $\beta>0$ and $\Delta\log(L)<0$ are likely objects that have lost mass and consequently are less luminous at the end of the interval. The only process that can give such result is the stripping of significant amount of stars during galaxy encounters. The galaxies that have $\beta<0$ and $\Delta\log(L)<0$ have decreased their luminosity at nearly constant mass, or their mass is moderately increasing (being $\Delta\log(\sigma>0$)), but the luminosity quenching is very big. The galaxies where $\beta>0$ and $\Delta\log(L)>0$ are still merging each other increasing both mass and luminosity. Finally the galaxies where $\beta<0$ and $\Delta\log(L)>0$ are those probably rich of gas that have lost stellar mass (being $\Delta\log(\sigma<0$)) during a close encounter. In this case, shocks might induce a burst of SF increasing the luminosity, but the total mass of the galaxies decreases.

According to the values of $\beta$ and $\Delta\log(L)$ each galaxy has its peculiar motion in the \Lsiga\ plane. The galaxies beloning to the upper part of the diagram with $\beta>0$ do move along the \Lsiga\ relation, while those with $\beta<0$ move perpendicularly to this relation.

\begin{figure}
	\centering
	\includegraphics[width=0.45\textwidth]{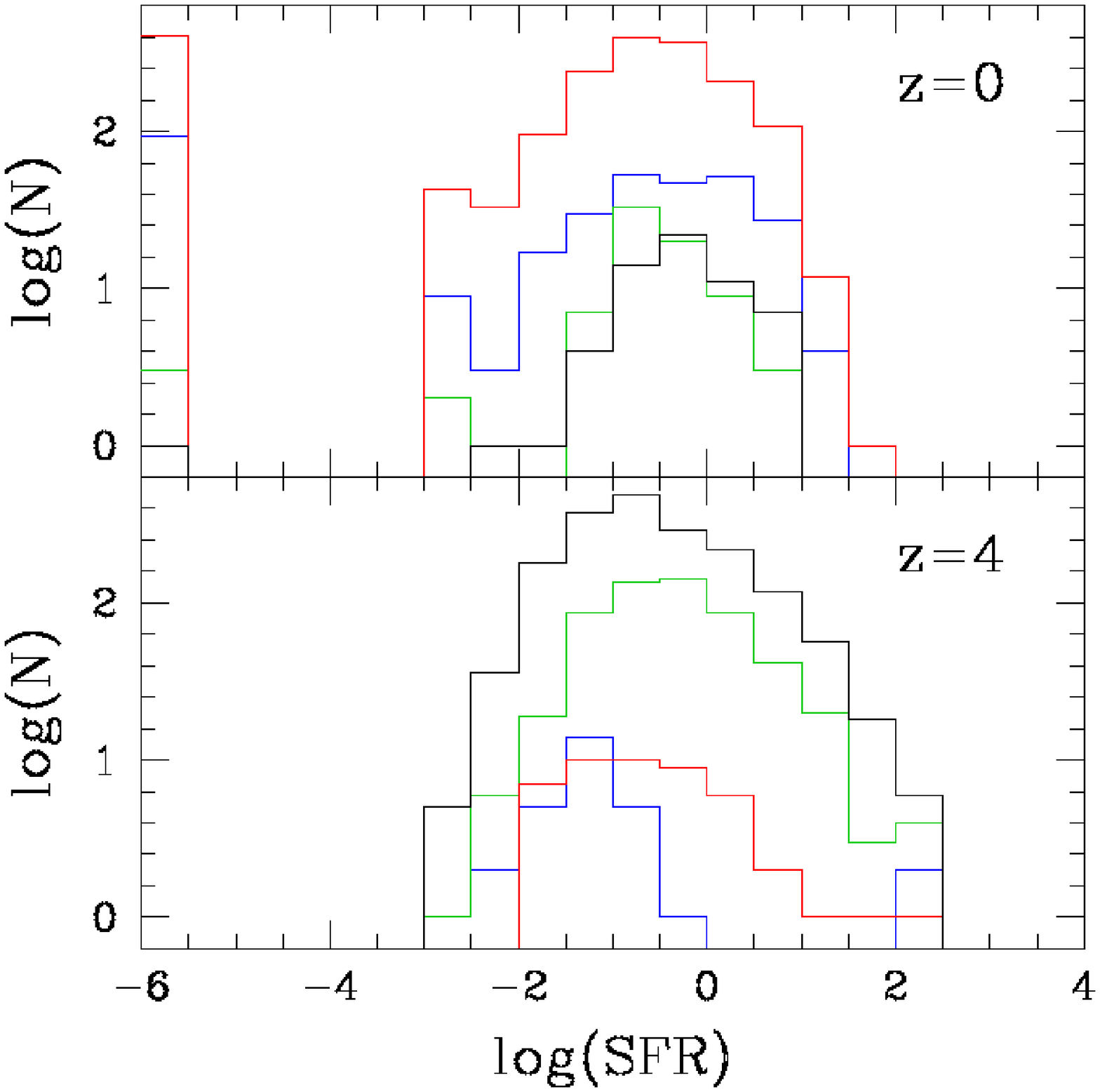}
	\caption{Upper panel: histogram of the SFR at $z=0$ for the galaxies belonging to the four parts of the $\beta-\Delta\log(L)$ diagram. Blue lines mark galaxies with $\Delta\log(L)<0$ and $\beta<0$, green lines mark those with $\Delta\log(L)>0$ and $\beta<0$, red lines those with $\Delta\log(L)<0$ and $\beta>0$, black lines those with $\Delta\log(L)>0$ and $\beta>0$. The galaxies that have $SFR=0$ have been assigned to the bin $10^{-6}$. Lower panel: the same histogram for the galaxies at $z=4$.}
	\label{fig:10}
\end{figure}

Figure \ref{fig:10} now shows the distribution of the SFR for the galaxies belonging to the different quadrants of Fig. \ref{fig:9} (left panel) at two redshift epochs: at $z=0$ (upper panel) and at $z=4$ (lower panel). 

We can see very clearly that at the two redshift epochs the situation is the opposite. At $z=0$ most of the galaxies are in the red and blue histograms, i.e. are galaxies that have lost significant amount of mass or are now in a quenched state ($SFR=0$). On the contrary at $z=4$ most of the galaxies are in the green and black histograms, i.e. are galaxies still in a merging phase or have formed new stars after a close encounter where they have lost part of their mass.

If we trust in simulations we can then deduce the following: \\
i) the galaxies that in the \Lsiga\ plane move perpendicularly to the FJ relation are those with $\beta<0$ that might be in a quenched state or might be experiencing a burst of SF after a close encounter where some of their mass is escaped.\\
ii) the galaxies that loose mass or acquire mass during encounters and merging move along the FJ relation.

In conclusion the present analysis has demonstrated that:\\
1. the distribution observed in the $I_e - R_e$ plane originate from the \Lsigb\ relation;\\
2. the distribution does not depend on the shape of the light profiles, but only on the history of mass assembly, mass removal, star formation and stellar evolution;\\
3. the distribution in the $I_e - R_e$ plane depends on the spectrum of values of $\beta$ (either positive and negative) that each galaxy can gain during its evolution;\\
4. all the scaling relations based on the effective radius $R_e$ can be interpreted as originating from the combination of the virial theorem and the \Lsigb\ relation.

It remains the problem of the linear relations: $M_V - n$, $M_V - \mu_0$ and \Lsig.
Why do not we see any curvature in these relations? Our explanation is that all these relations simply mirror  the virial relation, i.e. they are not real relations involving the luminosity of galaxies and its history. The real driver of these relations is the mass of the galaxies not the luminosity.
The luminosity enters in the relations only  when the half-luminosity radius $R_e$ is taken into account as a parameter.

The $M^* - n$ relation is linear in log units, as well as the \Lsig\ relation and the $M_V - \mu_0$ relation. $\mu_0$ is likely a parameter that does not hide in itself the history of a galaxy, made by mergers and star formation. 

The last consideration is that we should now admit that the \Lsigb\ relation has a solid empirical evidence. It is able to explain the distributions observed in the scaling relations involving luminosity and effective radius. This law must therefore be considered the 'true' FJ relation, i.e. the true relation connecting luminosity and velocity dispersion. Unfortunately, the connection of the parameters $\beta$ and $L'_0$ with the galaxy assembly and stars evolution is not easy, being such history a combination of different physical mechanisms operating at the same time. 

\begin{acknowledgements}
C.C. thanks the Department of Physics and Astronomy of the Padua University for the hospitality and
computational support.
\end{acknowledgements}

\end{document}